\documentclass[10pt,twocolumn,showpacs,amsmath,amssymb,superscriptaddress,bibnotes,floatfix,prb,aps]{revtex4-2}
\usepackage[dvipsnames]{xcolor}
\usepackage[T1]{fontenc}
\usepackage{amsmath}
\usepackage{braket}
\usepackage{simpler-wick}
\usepackage{tikz}
\usetikzlibrary{decorations.pathreplacing,decorations.markings, calc, shadows}
\tikzset{
  on each segment/.style={
    decorate,
    decoration={
      show path construction,
      moveto code={},
      lineto code={
        \path [#1]
        (\tikzinputsegmentfirst) -- (\tikzinputsegmentlast);
      },
      curveto code={
        \path [#1] (\tikzinputsegmentfirst)
        .. controls
        (\tikzinputsegmentsupporta) and (\tikzinputsegmentsupportb)
        ..
        (\tikzinputsegmentlast);
      },
      closepath code={
        \path [#1]
        (\tikzinputsegmentfirst) -- (\tikzinputsegmentlast);
      },
    },
  },
  mid arrow/.style={postaction={decorate,decoration={
        markings,
        mark=at position 0.53 with {\arrow[#1]{stealth}}
      }}},
}
\definecolor{viridisGreen}{RGB}{94, 201, 98}
\usepackage{graphicx}
\usepackage{hyperref}
\usepackage{placeins}

\renewcommand{\d}{\mathrm{d}}
\newcommand{\ham}{\mathcal{H}}

\newcommand{\cc}[1]{{#1}^{\dagger}}

\begin{document}
\preprint{APS/123-QED}

\title{Investigating Stark many-body localization with continuous unitary transformation flows}

\author{Jan-Niklas Herre}
 \email{jan.herre@rwth-aachen.de}
 \affiliation{Institute for Theory of Statistical Physics, RWTH Aachen University}

\author{Qiyu Liu}
 \affiliation{Technische Universität Braunschweig, Institut für Mathematische Physik, Mendelssohnstraße 3, 38106 Braunschweig, Germany}

 \author{Roman Rausch}
 \affiliation{Technische Universität Braunschweig, Institut für Mathematische Physik, Mendelssohnstraße 3, 38106 Braunschweig, Germany}

\author{Christoph Karrasch}
 \affiliation{Technische Universität Braunschweig, Institut für Mathematische Physik, Mendelssohnstraße 3, 38106 Braunschweig, Germany}
\author{Dante M.~Kennes}
 \email{dante.kennes@rwth-aachen.de}
 \affiliation{Institute for Theory of Statistical Physics, RWTH Aachen University}
 \affiliation{Max Planck Institute for the Structure and Dynamics of Matter,
Center for Free Electron Laser Science, Luruper Chaussee 149, 22761 Hamburg, Germany}
\date{\today}% It is always \today, today,

\begin{abstract}
We investigate the ergodicity-to-localization transition in interacting fermion systems subjected to a spatially uniform electric field.
For that we employ the recently proposed Tensorflow Equations (TFE), a type of continuous unitary flow equations. This enables us to
iteratively determine an approximate diagonal basis of the quantum many-body system. We present improvements to the method, which achieves good accuracy at small to intermediate interaction strengths, even in the absence of an electric field or disorder. Then, we examine two quantities that reveal the fate of Stark MBL in 1D and 2D. First, we
investigate the structure of the resulting basis to determine the crossover between ergodic and localized regimes with respect to electric field strength. Second, we simulate long-time dynamics at infinite temperature. Our results in 1D show a localization transition at non-zero field for finite interaction that vanishes with increasing system size leading to localization at infinitesimally small field even in the presence of interactions. In 2D we find less clear signatures of localization and strong finite size effects. 
We establish that the TFE work accurately up to intermediate times but cannot capture higher order effects in interaction strength that lead to delocalization at longer times in finite-size Stark MBL systems.
\end{abstract}

\maketitle
\section{Introduction}
One of the main challenges of condensed matter physics is to accurately and efficiently simulate interacting quantum many-body systems. Due to the exponentially growing size of the many-body Hilbert space this poses a formidable challenge. Still, over the years an array of approaches have been devised to tackle the quantum many-body problem beyond exact diagonalization. These range from the density matrix renormalization group (DMRG) \cite{White_1992,Schollw_ck_2011} including generalizations to tackle thermodynamics and dynamics \cite{Vidal2004, White_2004, Daley_2004,Feiguin_2005, Schmitteckert_2004, Vidal_2007, Kennes_2016}, which, however, is limited to low-entanglement, to more recent developments like Neural Quantum States (NQS) \cite{Carleo_2017}, with currently active debates about their benefits and limitations \cite{Deng_2017, Sharir_2022, Sun_2022, Luo_2023, Gutierrez2022, Passetti_2023}. Specifically for disordered systems progress in numerical techniques is urgently needed to accurately complement analytical treatments \cite{Goremykina_2019, Balasubramanian_2020, Monteiro_2021}. Since the discovery of Anderson localization \cite{Anderson_1958} the effect of random disorder on non-interacting particles has been well studied \cite{Evers_2008}. In the last two decades research focus shifted to the influence of interactions, and the stability of localization at non-zero temperature, turning this into a true many-body problem, called many-body localization (MBL) \cite{Fleishman_1980,Basko_2006,Gornyi_2005, _nidari__2008, Oganesyan_2007,Pal_2010, Ros_2015,Kj_ll_2014, Luitz_2015, Luitz_2017, Nandkishore_2015, Altman_2015, Serbyn_2015, Abanin_2019, Sierant2024}. To this day the fate of MBL in the thermodynamic limit and at infinite time remains debated, and people have mostly retreated to studying finite size systems at experimentally relevant time scales \cite{Morningstar_2022}.

Randomness is not necessary to obtain a many-body localized system. One can also obtain MBL by applying a quasi-periodic potential (QP-MBL) \cite{Iyer_2013, Lee_2017, Chandran_2017, Nag_2017, Khemani_2017, Lev_2017}, which is more attainable for experiments \cite{Schreiber_2015, L_schen_2017, Bordia_2017} and also removes the phenomenon of rare ergodic regions and avalanche effects \cite{De_Roeck_2017, Thiery_2018, Morningstar_2022}. However, it still breaks translation invariance and is analytically challenging. A new approach - in the same spirit as the logical step from Anderson localization to MBL - was considering systems subjected to a spatially uniform electric field. In absence of interaction this leads to \textit{Wannier-Stark} localization \cite{Wannier_1962}. Numerical evidence about the persistence of localization in presence of interactions, given a sufficiently strong electric field, was presented a few years ago \cite{Schulz_2019,van_Nieuwenburg_2019}. Notably, this model can be made translation invariant by employing a dynamic gauge, and applying a time-dependent vector potential, which is a type of dynamical localization \cite{Dunlap_1986} and draws a connection to the stability of MBL in periodically driven systems (Floquet-MBL) \cite{Ponte_2015, Abanin_2016, Bairey_2017}.
This showed that certain properties known from many-body localization could very well exist in clean finite-size systems, which would be of further advantage to experiment, eliminating the need of fine-tuning the potential. Subsequently, Stark MBL was shown to exist in several experiments, involving trapped ions \cite{Morong_2021}, cold atoms \cite{Scherg_2021} and superconducting qubits \cite{Wang_2021, Guo_2021}. 

In this work we employ a recently proposed numerical implementation of \textit{Continuous Unitary Transformations} \cite{Wegner_2000, Kehrein_2006}, an iterative method of separating energy scales and diagonalizing the Hamiltonian of a system in a renormalization group like flow, which has been applied to the MBL problem in the past \cite{Pekker_2017}. However, this method required either the full Hamiltonian matrix as input or the analytical setup of a large set of coupled ordinary differential equations (ODEs). This lead Refs.~\cite{Thomson_2024, Thomson2023a} to reformulate the ODEs in a numerically more general way, inspired from tensor contractions in tensor network codes, named \textit{Tensorflow Equations} (TFE). Here, we further develop the method and apply it to the Stark MBL problem.

The paper is set up as follows: In Section~\ref{section: model} we introduce the model in detail. We review TFE in Section~\ref{section: TFE} with more details in Appendix~\ref{section: Commutators} and \ref{section: Operator flow} and introduce how to compute dynamics in Section \ref{section: Dynamics}. An error analysis can be found in Appendix~\ref{section: energy error analysis}. Next, we present the results in Section~\ref{section: Results} which are twofold. First, we investigate the localization transition in Section~\ref{section: Transition} and second, we show long-time dynamics at infinite temperature in Section~\ref{section: dynamics results}, which includes a benchmark with simulations using the time dependent density matrix renormalization group (tDMRG). A comparison with the dynamics results obtained from the method introduced in Ref.~\cite{Thomson_2024} is given in Appendix~\ref{section: Dynamics comparison}. Finally, we comment on emerging delocalization behavior at long times of the Stark MBL model using the tDMRG data in Section~\ref{section: Delocalization Time}. We conclude this work with a discussion in Section~\ref{section: Discussion}.

\section{The Model}\label{section: model}

\begin{figure}[t]
  \begin{tikzpicture}
%     % Draw arrow
     \draw[-stealth, line width=1pt] (-1.7,-.8) -- (-1.6,-1.15)node[below]{{$x$}};
     \draw[-stealth, line width=1pt] (-1.7,-.812) -- (-1.72,-0.32)node[left]{{$h$}};
    \draw[-stealth, line width=1pt] (-1.7,-.8) -- (-1.33,-.74)node[above]{{$y$}};
    % Include graphic
   \node at (2.5,0) {\includegraphics[width=0.8\columnwidth]{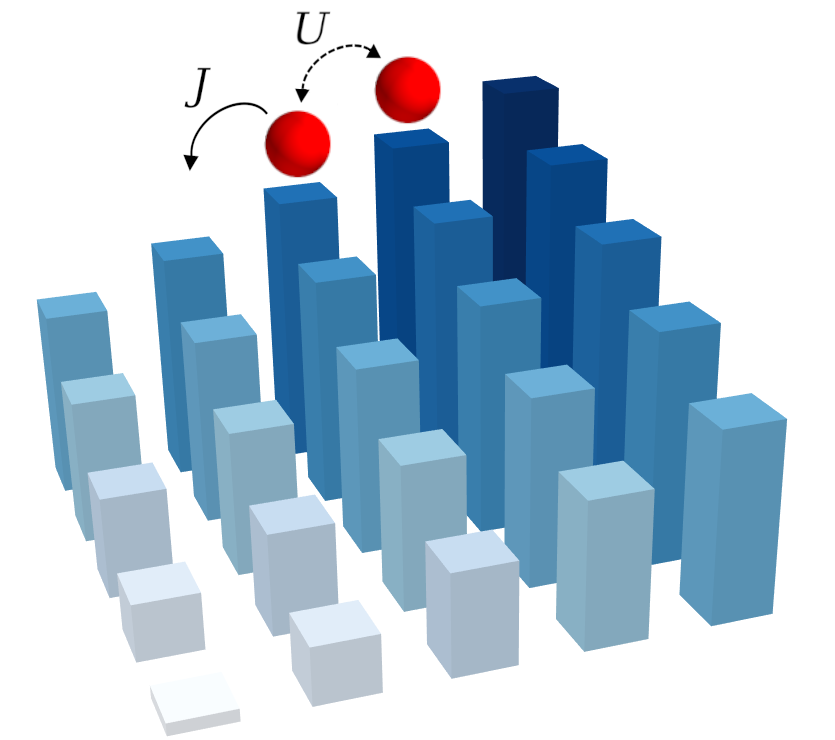}};
  \end{tikzpicture}
\caption{The Stark MBL system in two dimensions. Fermions hop with hopping amplitude $J$ and interact with strength $U$ on a tilted potential. Localization due to the potential competes with delocalization due to interactions.}
\label{fig: stark potential cartoon}
\end{figure}
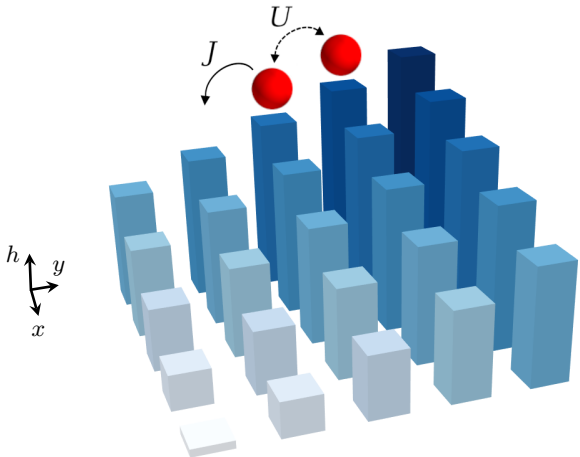
We consider a  system of spinless fermions with nearest neighbor interactions, that can be described by the Hamiltonian
\begin{equation}
    \ham = \sum_{j\in \Lambda} h_j :\cc{c}_j c_j: + \sum_{\langle jk\rangle} J_{ij} :\cc{c}_jc_k: + \sum_{\langle jk\rangle} U_{jk} :\cc{c}_jc_j\cc{c}_kc_k: \,, \label{eq:spinless_fermion_nn_interacting_model}
\end{equation}
with an external site-dependent potential $h_j$,
uniform nearest neighbor hopping $J_{jk}:=J=1$ and interaction $U_{jk}:=U$. $:...:$ indicates normal ordering. For example,
\begin{align}
    :\cc{c}_j c_k: &= \cc{c}_jc_k \,, \\
    :\cc{c}_j c_j\cc{c}_k c_k: &= -\cc{c}_j\cc{c}_kc_jc_k\,,
\end{align}
holds for normal ordering with respect to the vacuum state. Normal ordering is generally needed when computing flow equations, for it allows us to truncate the flow equations with a fixed lowest order without ambiguity, see Ref.~\cite{Kehrein_2006} for more details. We choose the lattice $\Lambda$ to be either a 1D chain or a simple cubic lattice with $L$ or $L^2$ sites respectively. In both cases we consider open boundary conditions and fix the particle number at half filling. In the one-dimensional case the Hamiltonian in Eq.~\eqref{eq:spinless_fermion_nn_interacting_model} can be mapped to the spin-$\frac{1}{2}$ XXZ-chain via a Jordan-Wigner transformation,
\begin{equation}
    \ham = \sum_{j\in\Lambda} J_{0}(S^{x}_{j}S^{x}_{j+1}+S^{y}_{j}S^{y}_{j+1}+\Delta S^{z}_{j}S^{z}_{j+1}) + h_{j}S^{z}_{j}
    \label{eq:spinful_XXZmodel}\,,
\end{equation}
with $J_{0} = 2J$, $J_{0}\Delta = U$. We focus on the case of \textit{Stark} MBL, in which application of a longitudinal electric field $\Vec{E}$ leads the system to become insulating even in presence of interactions \cite{Schulz_2019, Taylor_2020, Zisling_2022, Doggen_2022}. 
There is a subtlety in the choice of the exact form of the potential, regarding the presence of single-particle resonances, which is why often an additional small quadratic potential is introduced. However, since we cannot expect our perturbative method to resolve such resonances, we choose a solely linear onsite potential $h_j = -\gamma j$, where $\gamma$ denotes the spatially uniform electric field strength. In 2D, we apply the potential diagonally as $h_{xy} = -\gamma(x + y)$, such that $\Vec{E}=(\gamma, \gamma)$ and $|\Vec{E}|=\sqrt{2}\gamma$ is constant, see Fig.~\ref{fig: stark potential cartoon}. For all our simulations, the two-dimensional Hamiltonian is mapped to a 1D Hamiltonian with long-range hopping and interaction by snaking over the lattice which has been employed in different variations for tensor networks ~\cite{Cataldi_2021, Verstraete_2005}. We take the most simple approach, tracing the lattice row by row as shown in Ref.~\cite{Thomson_2024}.
\section{The Method} \label{section: Method}
\subsection{Continuous Unitary Transformations with Tensorflow Equations}\label{section: TFE}
To construct the \textit{Tensorflow Equations} (TFE) we start from Wegner's flow equation \cite{Wegner_2000}
\begin{equation}
    \frac{\d \ham(l)}{\d l} = \left[\ham(l), \eta(l)\right] \,,
    \label{flow equation}
\end{equation} which can be used to generate a set of
unitary equivalent Hamiltonians $\ham(l)$ starting from $\ham(0):=\ham$, using the anti-hermitian generator $\eta(l)$. We can verify the solution of Eq.~\eqref{flow equation} is given by:
\begin{equation}
    \ham(l) = U(l)\ham(0)U^{\dagger}(l)\;,
    \label{unitary transformation}
\end{equation} where $U(l)$ has the form: $U(l) = \mathcal{T}_l \text{exp} \big( - \int_{0}^{l} \eta (l') dl' \big)$. By $\mathcal{T}_l$ we denote the time-ordering operator with respect to the artificial flow parameter $l$. The goal is to find a generator for the unitary transformation $U(l)$ that diagonalizes $\ham(l)$ in the limit of $l\rightarrow \infty$.
We follow the implementation of the TFE introduced in Refs.~\cite{Thomson_2024, Thomson2023a}, that manages to reduce the analytical work of setting up explicit flow equations for the couplings (see for example Ref.~\cite{Kehrein_2006}) by borrowing ideas from general tensor-contraction techniques. To represent the running Hamiltonian during the flow we rewrite our Hamiltonian defined in Eq~\eqref{eq:spinless_fermion_nn_interacting_model} in its most general form in terms of fermionic operator strings:
\begin{align}
    &\ham(l)= \ham^{(2)}(l) +  \ham^{(4)}(l) + \dots \\
    =&  \sum_{\alpha\beta} \ham^{(2)}_{\alpha\beta}(l) :c^{\dagger}_{\alpha}c_{\beta}: + \sum_{\alpha\beta\gamma\delta} \ham^{(4)}_{\alpha\beta\gamma\delta}(l) :c^{\dagger}_{\alpha}c_{\beta}c^{\dagger}_{\gamma}c_{\delta}: + \dots
\label{eq:generic_hamiltonian}
\end{align}
The kinetic and interaction terms are now encoded in the coefficient tensors with two and four legs, while allowing for terms up to infinite order that are generated during the flow.
Let $\ham_{\text{0}}(l)$ ($\ham_{\text{off}}(l)$) be the (off-)diagonal part of $\ham(l)$. The canonical generator introduced by Wegner \cite{Wegner_2000} is then given by
\begin{align}
    \eta_{W}(l) &:=\left[\ham_{0}(l), \ham_{\text{off}} (l)\right] = \left[\ham(l), \ham_{\text{off}}(l) \right]  \nonumber\\
    &\;= \underbrace{\left[\ham_{}^{(2)}(l), \ham_{\text{off}}^{(2)} (l)\right]}_{=:\eta_{W}^{(2)}(l)}  \nonumber\\
    &\;\; + \underbrace{\left[\ham_{}^{(2)}(l), \ham_{\text{off}}^{(4)}(l) \right] + \left[\ham_{}^{(4)}(l), \ham_{\text{off}}^{(2)}(l) \right]}_{=:\eta_{W}^{(4)}(l)} + \dots
    \label{def: Wegner Generator by order}
\end{align} 
If we omit terms containing more than $4$ fermionic operators, i.e. everything beyond two-particle interaction, the flow equations become a closed system. This truncation is an accurate approximation for small interaction strengths and accuracy can be systematically improved by including higher order terms. We will not include terms beyond such two-particle terms for simplicity and increased performance. Details about how to compute the appearing commutators up to this order can be found in Appendix~\ref{section: Commutators}. The completed flow gives an approximate solution to the eigenvalue problem posed by $\ham$.
\paragraph*{Choice of Generators ---}
The canonical generator provides stable convergence of the flow for most problems even though it has the downside of generating more long-ranged off-diagonal couplings in the early stage of the flow.  Employing the Wegner generator, couplings are suppressed rapidly when they are associated with two diagonal elements with large energetic difference, which also means it becomes inefficient when dealing with degeneracies. At zero external potential it breaks down completely. Fortunately, this limitation can be mitigated through a combination with the Toda-Mielke generator \cite{Mielke_1998}
\begin{equation}
    \lambda_{\alpha\beta}(l) = \text{sgn}(\beta-\alpha)\ham^{(2)}_{\alpha\beta}(l)
    \label{def: Toda Generator}\,,
\end{equation} which allows to lift degeneracies while preserving the band structure of the initial matrix.
Ref.~\cite{Thomson_2024} introduced this combination of two generators as a ''scrambling'' transformation. They invoked a full unitary transformation with a variant of the generator that only triggers if single particle states are near degenerate. The generator can be written as 
\begin{equation}
    \lambda_{\alpha \beta}(l) = \begin{cases}
       \text{sgn}(\beta - \alpha)\ham^{(2)}_{\alpha\beta}(l) \,, \;\, &\text{if } \ham^{(2)}_{\alpha\beta}(l) \geq \delta h \,, \\ 
       0 \,, \;\, &\text{else}\,.
    \end{cases} 
    \label{eq:Scrambling Generator}
\end{equation}
Here we defined $\delta h = \varepsilon |\ham^{(2)}_{\alpha\alpha}(l) - \ham^{(2)}_{\beta\beta}(l)|$
where $\varepsilon$ denotes a non-zero threshold. This generator is used to perform a full unitary transformation $\ham \to \cc{\mathcal{S}}\ham\mathcal{S}$ where the infinitesimal transformation is given by $\d S = \mathrm{exp}(-\lambda \d l)$. The unitary $\mathcal{S}$ is applied as long as near degenerate states below the defined threshold are encountered. Next, a canonical flow with Eq.~\eqref{def: Wegner Generator by order} is employed to diagonalize the Hamiltonian. Since the step of applying several scrambling transformations $\mathcal{S}$ can be rather expensive and is only justified for small external potential, we additionally investigated the benefits of a combination of canonical Wegner generator and Toda-Mielke generator (Eq.~\eqref{def: Toda Generator}) in one flow. I.e., instead of a full scrambling transformation we ''enhance'' the Wegner generator by adding Eqs.~\eqref{def: Toda Generator} or \eqref{eq:Scrambling Generator} to it at each infinitesimal flow step. 
We compared the performance of the different generators in Appendix~\ref{section: energy error analysis} and found that the full scrambling transformation followed by a canonical flow provides the best results, diagonalizing Hamiltonians with zero external potential. This setup is used to obtain all results in Section~\ref{section: Results}.
\paragraph*{Implementation ---}
The implementation of the flow equations follows the one in Ref.~\cite{Thomson_2024}. We implement the flow equations in python using \textit{JAX} just-in-time compilation to GPU. The ODEs are solved with the \textit{scipy} implementation of a 4/5 Runge-Kutta stepper. The implementation of the dynamics is also done with \textit{JAX} and runs GPU compiled (see Section~\ref{section: Dynamics}).
\paragraph*{Convergence ---}
We run the flow until $l=1000$ or $\mathrm{max}|\ham^{(2)}_{\mathrm{off}}| < 10^{-6}$ and $\mathrm{max}|\ham^{(4)}_{\mathrm{off}}| < 10^{-4}$. The second condition is harder to meet as the flow of the interacting elements converges much slower. So, additionally, given $\mathrm{max}|\ham^{(4)}_{\mathrm{off}}|\sim 10^{-n}$, we stop the flow if for the last 100 iterations there was no improvement of at least $\sim 10^{-n}$ in magnitude. An analysis of the convergence error is shown in Appendix~\ref{section: energy error analysis} for different choices of generators.
\paragraph*{Benchmarking ---}
We benchmark our TFE implementation against two methods: exact diagonalization (ED) and the time dependent density matrix renormalization group (tDMRG) implemented with matrix product states (MPS) \cite{Haegeman2011}.
 For ED comparison we use the \textit{QuSpin}-package \cite{Weinberg_2017, Weinberg_2019} for construction of the Hamiltonian and solve the eigenvalue problem as well as the dynamics on GPU using matrix routines from the \textit{cupy} library. We analysed the error in the energies resulting from TFE with respect to ED in Appendix~\ref{section: energy error analysis} and found it to scale qualitatively with $U/\gamma$.
 In order to test the validity of the TFE for larger systems than accessible by ED (see Fig.~\ref{fig:stark_1d_2d_dyn_corr}), we use tDMRG. This is only done in 1D. Here we construct the spin-$\frac{1}{2}$ XXZ-chain as an MPS, using the mapping in Eq.~\eqref{eq:spinful_XXZmodel}. We compute the dynamical spin autocorrelation $C(t) = \langle S_z(t) S_z(0)\rangle$ at infinite temperature using purification \cite{Verstraete2004}. 
TFE observables are computed within the half-filling sector and, correspondingly, this requires that the $z$-component of the total spin of the physical sites needs to be fixed to zero. As introduced in Ref.~\cite{Nocera2016}, the $z$-component of the total spin of both physical and auxiliary sites can be conserved and fixed to zero at the same time in the purified setup. 
\subsection{Computing Dynamics}\label{section: Dynamics}
In the TFE framework observables can be computed by letting a chosen operator flow alongside the Hamiltonian, finding its representation in the diagonal basis of $\ham$ (see Appendix~\ref{section: Operator flow}).
We calculate the dynamical autocorrelation function with the number operator $n_i$ at lattice site $i$ at infinite temperature
\begin{equation}
    C(t) = \left\langle \left( n_i(t) - \frac{1}{2}\right) \left( n_i(0) - \frac{1}{2}\right) \right\rangle \,. \label{eq:dynamical_autocorr}
\end{equation}
We fix $i=L/2$, the center of the system, throughout this work.
This quantity is related to the experimentally observable occupation imbalance \cite{Schreiber_2015, Hess_2017} and a non-zero value at long times is a signature
for the system keeping memory of its initial configuration and being an insulator. A drop of $C(t)\to0$ indicates thermalization and metallic behavior.
Given an (approximately) diagonalized Hamiltonian, we can compute Eq.~\eqref{eq:dynamical_autocorr} by tracing over all eigenstates $\{\ket{j}\}$, where we restrict ourselves to the half-filling sector. In the energy eigenbasis time evolution amounts to a phase and, by inserting an identity, the correlator can be written as
\begin{align}
\langle n_i(t)n_i(0) \rangle &= \langle e^{i\ham t}n_ie^{-i\ham t}n_i\rangle \\
    &=  \frac{1}{\mathcal{Z}}\sum_{j} \bra{j}e^{i\ham t}n_ie^{-i\ham t}n_i\ket{j}\nonumber\\
    &=  \frac{1}{\mathcal{Z}}\sum_{j,k} \bra{j}e^{iE_{j}t}n_ie^{-iE_{k}t}\ket{k}\bra{k}n_i\ket{j}\nonumber\\
    &= \frac{1}{\mathcal{Z}}\sum_{j,k} e^{i(E_{j} - E_{k})t} |\langle j|n_i|k\rangle|^2\; ,
    \label{eq: Dynamical Autocorrelation from Eigenenergies}
\end{align}
where $\mathcal{Z} = \sum_{j} e^{-E_{j} /T}$ is the partition function with $\mathcal{Z}\to\mathcal{N}$, the Hilbert space size for $1/T \to 0$ and $n_{i} \equiv n_{i}(0)$. In that way we can calculate the correlation function in Eq.~\eqref{eq:dynamical_autocorr} both in ED and the TFE, since TFE give us access to the energies (up to truncation error) and eigenstates are simple product states. 
\paragraph*{Computing matrix elements ---}{
We need to compute all matrix elements $\bra{j} n_i \ket{k}$ by applying $n_i$ to the eigenstates, represented by binary strings $\ket{j} \equiv \ket{011...10}$. Since we work in the diagonal basis and the operator is given by the expansion in fermionic operator strings in Eq.~\eqref{eq: Flowed Number operator} (acting on fermions \textit{in the diagonal basis}) we calculate
\begin{align}
    \bra{j} n_i \ket{k} =& \sum_{\alpha\beta} A_{\alpha\beta} \bra{j} \cc{c}_{\alpha}c_{\beta}\ket{k}\nonumber \\ 
    &- \sum_{\alpha\beta\gamma\delta} B_{\alpha\beta\gamma\delta}  \bra{j}\cc{c}_{\alpha}\cc{c}_{\gamma}c_{\beta}c_{\delta}\ket{k}\, ,
    \label{eq: TFE matrix elements}
\end{align}
where we implicitly assumed normal ordering and $A$ and $B$ are the coupling coefficients resulting from the flow of $n_i$. The two-body term can be simplified by applying Wick contractions with respect to the vacuum state. The Wick contraction can be written as \cite{Thomson_2024}
\begin{align}
    \bra{j}\cc{c}_{\alpha}\cc{c}_{\gamma}c_{\beta}c_{\delta}\ket{k}  =&\wick{
    \bra{j} \c1{c}^{\dagger}_{\alpha} \c2{c}^{\dagger}_{\gamma} \c2{c_{\beta}} \c1{c_{\delta}} \ket{k} + \bra{j} \c1{c}^{\dagger}_{\alpha} \c2{c}^{\dagger}_{\gamma} \c1{c_{\beta}} \c2{c_{\delta}} \ket{k}}\nonumber \\
    =& \bra{j}\cc{c}_{\alpha}c_{\delta}\ket{k}\cdot\bra{j}\cc{c}_{\gamma}c_{\beta}\ket{k} \nonumber \\
    &- \bra{j}\cc{c}_{\alpha}c_{\beta}\ket{k}\cdot\bra{j}\cc{c}_{\gamma}c_{\delta}\ket{k} \, .
\end{align}
The evaluation of the resulting hopping terms is achieved by logically comparing the two binary strings representing the states. As an example, let us assume a system of $4$ sites and two states  $\ket{j} = \ket{0101}$ and $\ket{k} = \ket{1001}$. Evaluating the matrix elements amounts to applying one specific hopping operator at a time, like $\bra{j}\cc{c}_1 c_0\ket{k} = 1$. Only those states contribute that are reachable from a given state with a single particle hop. Considering only valid hoppings from $n_1$ occupied to $n_0$ unoccupied sites, given a system with in total $N = n_0 + n_1$ sites, there exist $n_{0} \cdot n_{1}$ reachable states.
In practice, we only include eigenstates at half filling, therefore we get $N^2/4$ reachable states (the case of maximal connectivity). For systems larger than accessible with ED, it is not feasible to trace over all eigenstates. Instead, we take the average of the expectation value of randomly drawn samples $\{\ket{j'}\}$ of size $M$. Since the second set of states $\ket{k'}$ only needs to contain those states that are connected by a single particle hop, the algorithm remains polynomial in scaling. 
}
\paragraph*{Error Estimate ---}{
There are two sources of errors affecting the dynamics simulation. The first is introduced by the sampling mechanism. A similar approach is called dynamical quantum typicality, where one evaluates the correlation function with respect to a random superposition of states drawn from the Haar measure \cite{Heitmann_2020}. There, the error scales as $1/\sqrt{\mathcal{N}_{eff}}$, where $\mathcal{N}_{eff}$ is the effective Hilbert space dimension, including all thermally occupied states. At infinite temperature and superpositioning \textit{all} eigenstates, this becomes exact. By identifying $\mathcal{N}_{eff} = M$ we find the same scaling for our method $1/\sqrt{M}$. In practice, we find it sufficient to choose $M=512$.\\
The second source of error stems from the eigenenergy error introduced by the TFE (see Section~\ref{section: energy error analysis}). Let us define the maximal absolute error as $\delta_E = \underset{j}{\mathrm{max}}(E^{(j)}_{ED}\varepsilon^{(j)}_{ED})=\underset{j}{\mathrm{max}}(E^{(j)}_{TFE} - E^{(j)}_{ED})$. We can then write the error in the dynamics of one eigenstate 
\begin{align}
    \epsilon_{\mathrm{dyn}} &= \frac{\left| (e^{iE^{(j)}_{TFE}t} - e^{iE^{(j)}_{ED}t}) \langle n_i \rangle \right|}{\left|e^{iE^{(j)}_{ED}t}\langle n_i \rangle \right|} \nonumber\\
                          &\leq\left| (e^{i(E^{(j)}_{TFE} - E^{(j)}_{ED})t} - 1)\right| \nonumber \\
                          &\leq \frac{1}{2}\delta^2_Et^2 \,.
\end{align}
}
We found that this method of computing dynamics performs significantly better than solving the Heisenberg equation of motion, which was done in Ref.~\cite{Thomson_2024}, see Appendix~\ref{section: Dynamics comparison} for a comparison.
\section{Results}\label{section: Results}
\subsection{Localization Transition in 1D and 2D} \label{section: Transition}

\begin{figure}[htbp]
\includegraphics[width=1\columnwidth]{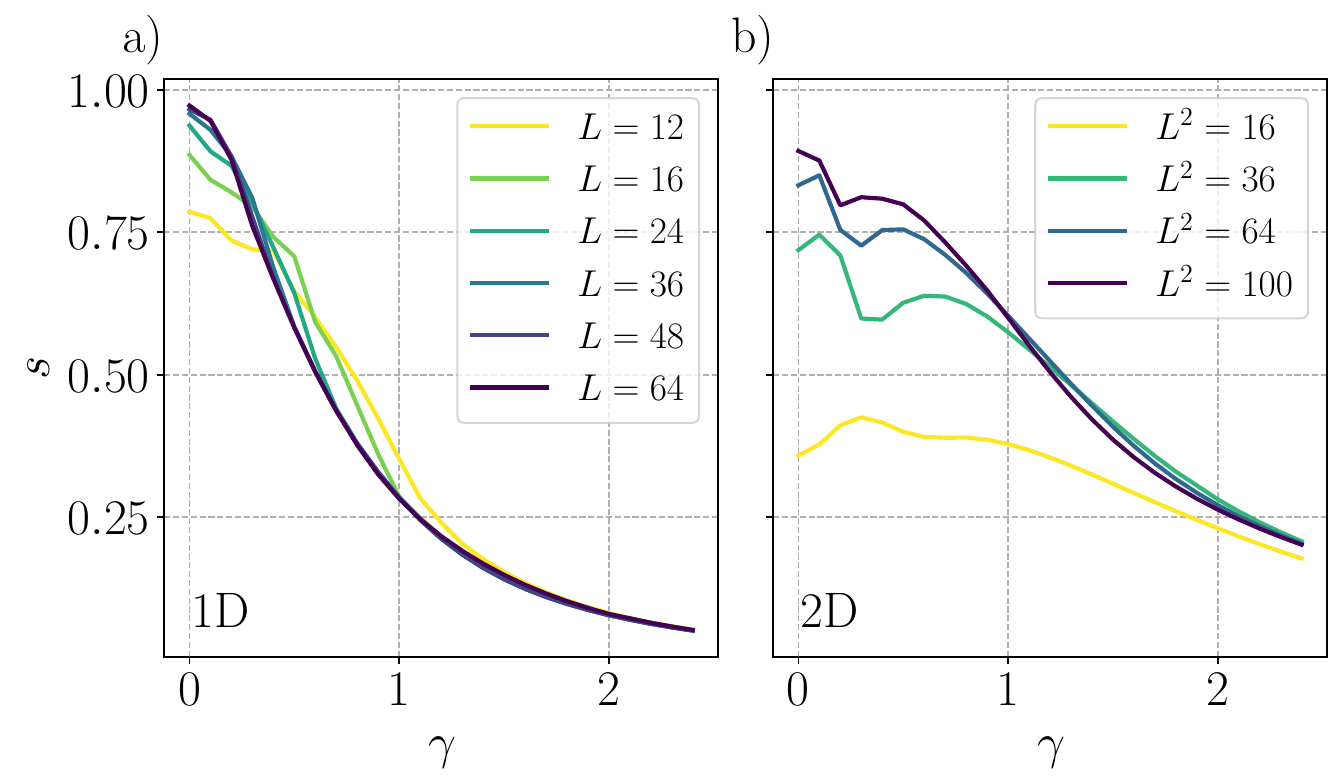}
\centering
\caption{a) The support of the number operator in the 1D Stark MBL system with $U=1.0$. A transition is visible at $\gamma_c\approx 0.7$ for small system sizes accessible by ED. For larger systems the transition drifts below $\gamma_c \approx 0.5$. b) Same results for a 2D Stark MBL system. The transition is at $\gamma_c \approx 1.3$ and a stronger finite size drift is visible for to small systems. 
}
\label{fig:stark_1d_transition}
\end{figure}
To analyse the localization transition with TFE, we refrain from using the established method of spectral statistics, because computing the mean adjacent energy gap \cite{Luitz_2015, Tikhonov_2018, Zisling_2022} or Heisenberg and Thouless timescales \cite{Suntajs_2020} requires accurate knowledge of the energies, that by now only ED can provide. We instead look at the structure of integrals of motion (IOM) in the definition of Refs.~\cite{Chandran_2015, Singh_2021}. An IOM can be defined via the long-time average of a chosen local operator $\mathcal{O}$, which is 
diagonal in the energy eigenbasis
\begin{align}
    \overline{\mathcal{O}} &= \lim_{t\to\infty}\frac{1}{t}\int_0^{t} \d t' \mathcal{O}(t') \\
   &= \sum_{j} \bra{j} \mathcal{O}\ket{j} \ket{j}\bra{j} \, .
\end{align}
We choose the local operator to be the number operator $\mathcal{O} \equiv n_i = \cc{c}_i c_i$.
By a suitable unitary transformation $\mathcal{U}$ we can relate the number operator to a maximally localized integral of motion (LIOM)
\begin{align}
    \tilde{n}_i = \mathcal{U} n_i \cc{\mathcal{U}}\, , \;\; \cc{\tilde{c}}_i = \mathcal{U} \cc{c}_i \cc{\mathcal{U}}\, ,
\end{align}
and expand the operator in terms of these new operators, which are diagonal in the energy eigenbasis.
\begin{equation}
    n_i = \sum_{N, \{\boldsymbol{\alpha}\}_N} \mathcal{C}^{(N)}_{\{\boldsymbol{\alpha}\}} \cc{\tilde{c}}_{\alpha_1} \tilde{c}_{\alpha_2}...\cc{\tilde{c}}_{\alpha_{N-1}} \tilde{c}_{\alpha_{N}} \, , \label{eq:LIOM_expansion}
\end{equation}
where $\{\boldsymbol{\alpha}\}_N = \{\alpha_1,\alpha_2,...,\alpha_N\}$ is a set of indices with a length $N$ corresponding to the expansion order.
By relaxing the requirement that the unitary transformation $U$ needs to be quasi-local, we recognize that letting $n_i$ flow with the TFE will lead to such a representation. The localization of the resulting LIOMs indicates overall localization of the system.
The flow results in the number operator written in terms of the tensor coefficients $\mathcal{C}^{(2)}_{\{\boldsymbol{\alpha}\}} =A_{\alpha\beta}$ and $\mathcal{C}^{(4)}_{\{\boldsymbol{\alpha}\}} = B_{\alpha\beta\gamma\delta}$ as in Eq.~\eqref{eq: TFE matrix elements}, implying a truncation of the expansion in Eq.~\eqref{eq:LIOM_expansion} at $4$th order. We can define a support quantity $s$
as the sum of the coefficients:
\begin{align}
s &= \frac{\sum_{\alpha\beta\gamma\delta} B_{\alpha\beta\gamma\delta}^2}{\sum_{\alpha\beta} A_{\alpha\beta}^2 + \sum_{\alpha\beta\gamma\delta} B_{\alpha\beta\gamma\delta}^2}  \,.
    \label{eq:liom space support}
\end{align}
In the ergodic regime the expansion is dominated by two-point interaction terms, $s\to 1$. On the other hand,
the interaction terms decay in the localized phase and the dominant contributing terms are the single particle terms ($s \to 0$).
Ref.~\cite{Bertoni_2024} introduced an equivalent quantity, basing it upon the so-called ''real-space'' support introduced in Ref.~\cite{Thomson2023a} and applying it to 1D Stark MBl systems with an earlier variant of TFE. 
We find that it is not necessary for the TFE flow to converge to particularly high precision for $s$ to display the characteristic structure of the LIOMs, since it only requires the global structure to be captured accurately, and we can study systems with $U=1.0$, on the edge of the TFE's capabilities.\\
In Fig.~\ref{fig:stark_1d_transition} we show the support $s$ as a function of the onsite potential $\gamma$ for different system sizes. We extract the (moving) critical potential strength $\gamma_c$, that signifies the finite size crossover between the delocalized and localized regimes, from the crossing point of the curves belonging to different system sizes. In 1D, finite size effects are weak and for small systems we estimate a (almost stationary) crossing point $\gamma_c \approx 0.7$, which agrees with results that can be obtained from ED spectral statistics \cite{Zisling_2022} and experiments with trapped ions \cite{Morong_2021}. However, for systems larger than accessible by ED, there is a visible drift of the crossing point towards smaller external potential, agreeing with the results from an earlier version of TFE \cite{Bertoni_2024}. An extrapolation to infinite system sizes could very well lead to $\gamma_c \to 0$, supporting the claims of \cite{Doggen_2021} that there is no true ergodic regime for $\gamma>0$ in the thermodynamic limit. 
This behavior is very different from the one found in ordinary MBL systems, see for example the analysis of wavefunction statistics in Refs.~\cite{Tikhonov_2021,Herre_2023}, spectral statistics in Ref.~\cite{Tikhonov_2018}, or Refs.~\cite{Iyer_2013,Zhang_2018,Aramthottil_2021} for scaling analyses of QP-MBL systems. 
Additionally, we need to consider that only the $s\to0$ limit can ensure true localization, as the problem then reduces to a non-interacting Stark problem, which is only achieved in the large $L$ limit or for very strong $\gamma$. Any $s>0$ allows for two-particle effects which could facilitate delocalization at infinite time in finite size systems either by themselves or by generating higher order effects that are not included in the truncated flow equations \cite{Kloss2023, Zisling_2022}. In 2D, finite size effects are much stronger and the crossing point is only visible beyond system sizes accessible by ED. The decrease to $s\to 0$ that signifies full localization is slower than in 1D, for larger systems we find a crossing point at $\gamma_c \approx 1.3$, slowely drifting to smaller $\gamma$. However, from the system sizes accessible by TFE, a statement about the thermodynamic limit would still be speculative. Our findings, which indicate localization even in 2D, do not contradict results about subdiffusive transport in 2D lattice systems with strong tilted potentials in theory \cite{Zhang_2020}, DMRG studies \cite{Doggen_2022} and experiment \cite{Guardado_Sanchez_2020}. This is due to the fact that we chose the most transport-prohibitive direction for the potential along the diagonal of the 2D grid, while the subdiffusive regime is found by applying the tilt along one of the axes, decoupling transport along the other axis from the potential. It would be important to conduct experiments with our setup to confirm these findings.
\subsection{Dynamics at Long Times}\label{section: dynamics results}
\begin{figure}[htbp]
\includegraphics[width=1\linewidth]{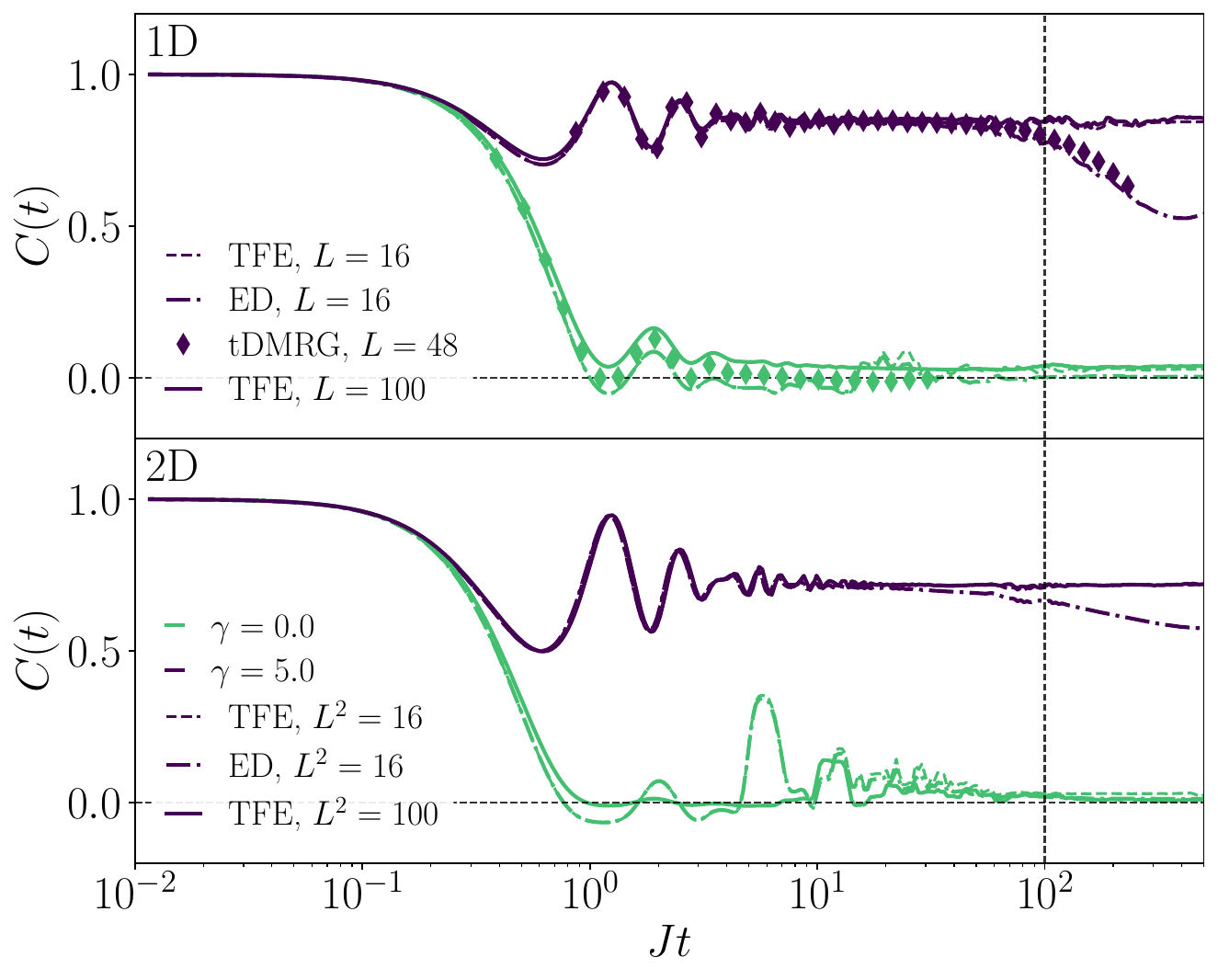}
\caption{The dynamical autocorrelation for the Stark MBL model in one (top panel) and two dimensions (bottom panel). Interaction strength is set to $U=0.1$. We expect accurate results until times $t\lesssim U^{-2}=100$ (dashed vertical line). Different system sizes are shown, together with ED results for a small system ($L=16$) and tDMRG results for $L=48$. tDMRG is only shown at late times for visibility. Data is averaged with a moving average due to strong oscillations (see Fig.~\ref{fig:stark_1d_dyn_corr_comparison} for raw data).}
\label{fig:stark_1d_2d_dyn_corr}
\end{figure}

\begin{figure}[htbp]
\includegraphics[width=1\linewidth]{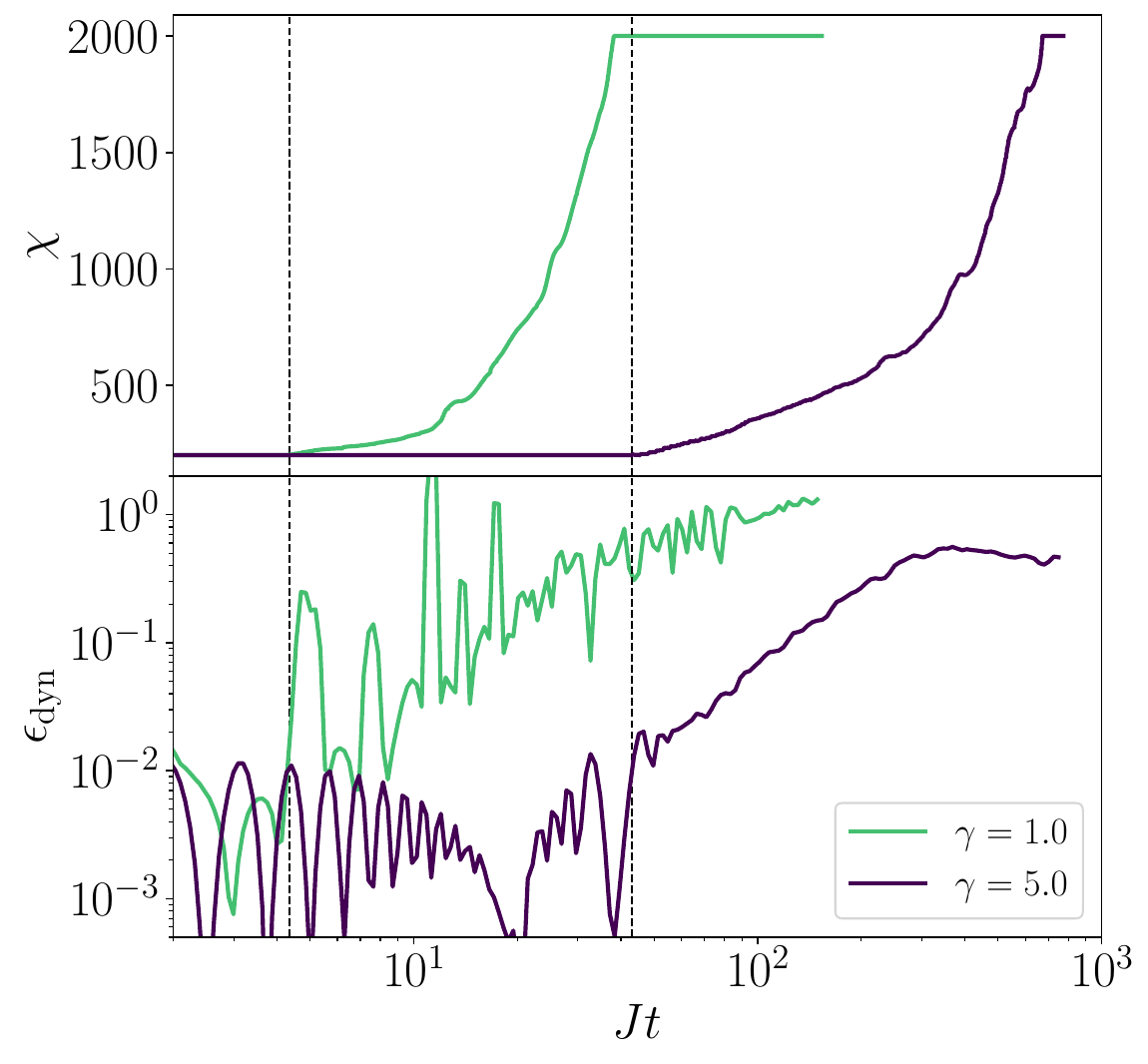}
\caption{Increase of bond dimension $\chi$ in tDMRG simulations of the dynamical autocorrelation at infinite temperature vs.~the dynamical error $\epsilon_{\mathrm{dyn}}$ accrued in the TFE simulation (with respect to tDMRG) for $L=24$, all other parameters are the same as in Fig.~\ref{fig:stark_1d_2d_dyn_corr}. The times at which the error and bond dimension start to increase coincides for both $\gamma=1.0$ and $\gamma=5.0$ (vertical dashed lines).}
\label{fig:stark_1d_dyn_corr_tdvp_err}
\end{figure}
  In Fig.~\ref{fig:stark_1d_2d_dyn_corr} we show the time evolution of the autocorrelation function (Eq.~\eqref{eq:dynamical_autocorr}) for both 1D and 2D Stark MBL systems and different potential strengths with ED results as comparison. The data is obtained for a logarithmic time grid from $t=0.01$ to $t=10^5$ with $5000$ steps. The autocorrelation function oscillates strongly, with fast oscillations dampened by interactions, which was also reported in Ref.~\cite{Doggen_2022}. For visibility and to focus on the global behavior we applied a moving average given by the unweighted mean of the previous 50 time steps. In 1D we also show tDMRG results for the spin correlation function $C(t) = \langle S_z(t)S_z(0)\rangle$ of the model in Eq.~\eqref{eq:spinful_XXZmodel} as a benchmark. The data is obtained with a cutoff bond dimension $\chi_{\mathrm{max}}=2000$ and a fixed time step size of $\Delta t=0.05$. The moving average interval is adapted accordingly and only data for late times is shown for visibility (raw data is shown in Fig.~\ref{fig:stark_1d_dyn_corr_comparison}).\\
In the 4th order approximation, neglected terms limit the energy resolution to the order $\mathcal{O}(U^2)$. Consequently, dynamics can be simulated accurately up to times $t\sim 1/U^2$. 
The threshold is marked as a vertical dashed line and, indeed, this is the time at which qualitative deviations from the ED results start to occur. For random and quasi-periodically disordered systems, one finds that the TFE stay accurate for longer times \cite{Thomson_2024}, since averaging over configurations can average out the error. Additionally, random and QP-MBL models do not display any emergent behavior at those time scales. By looking at the Stark MBL model with intricate behavior at long times - a finite size delocalization effect already presented in Refs.~\cite{Zisling_2022,Kloss2023} - we observe a breakdown of accuracy at the expected intermediate time scale.\\
To get a better understanding of the breakdown we compare the TFE dynamical error to the bond dimension growth in the tDMRG simulation. This is shown in Fig.~\ref{fig:stark_1d_dyn_corr_tdvp_err}. While tDMRG is limited by bond dimension growth, especially in the delocalized regime, it stays accurate until $\chi_{\mathrm{max}}$ is reached. The TFE, on the other hand, break down as soon as higher order many-body effects become important which coincides with the time where bond dimension starts to increase. Still, we manage to go far beyond the system sizes accessible by ED and, in the delocalized regime, even tDMRG and simulate the dynamics accurately up to intermediate times. Only at late times and in the localized regime, tDMRG starts to outperform TFE in both accessible system size and accuracy.\\
The start of both the TFE error increase and tDMRG bond dimension growth coincide with the onset of delocalization. This leads us to conclude that this effect is of higher order than the TFE truncation captures and is accompanied by increasing entanglement entropy. One possible explanation for this long term delocalization time could be that localization is driven by Hilbert space fragmentation, which leads to a strong dependency of the dynamics on initial states \cite{Doggen_2022}. At late times, higher order interaction processes start to hybridize delocalized and localized states in the infinite temperature dynamics, leading to thermalization. This could also explain why the infinite temperature autocorrelation function shows this phenomenon (see also Ref.~\cite{Zisling_2022}), while imbalance simulations of specific initial states, like charge density waves, do not \cite{Schulz_2019}. Therefore, our results also do not contradict findings of localization in quench experiments \cite{Wang_2021, Morong_2021}.

\subsection{Scaling of the Delocalization Time} \label{section: Delocalization Time}
\begin{figure}[htbp]
\includegraphics[width=1\linewidth]{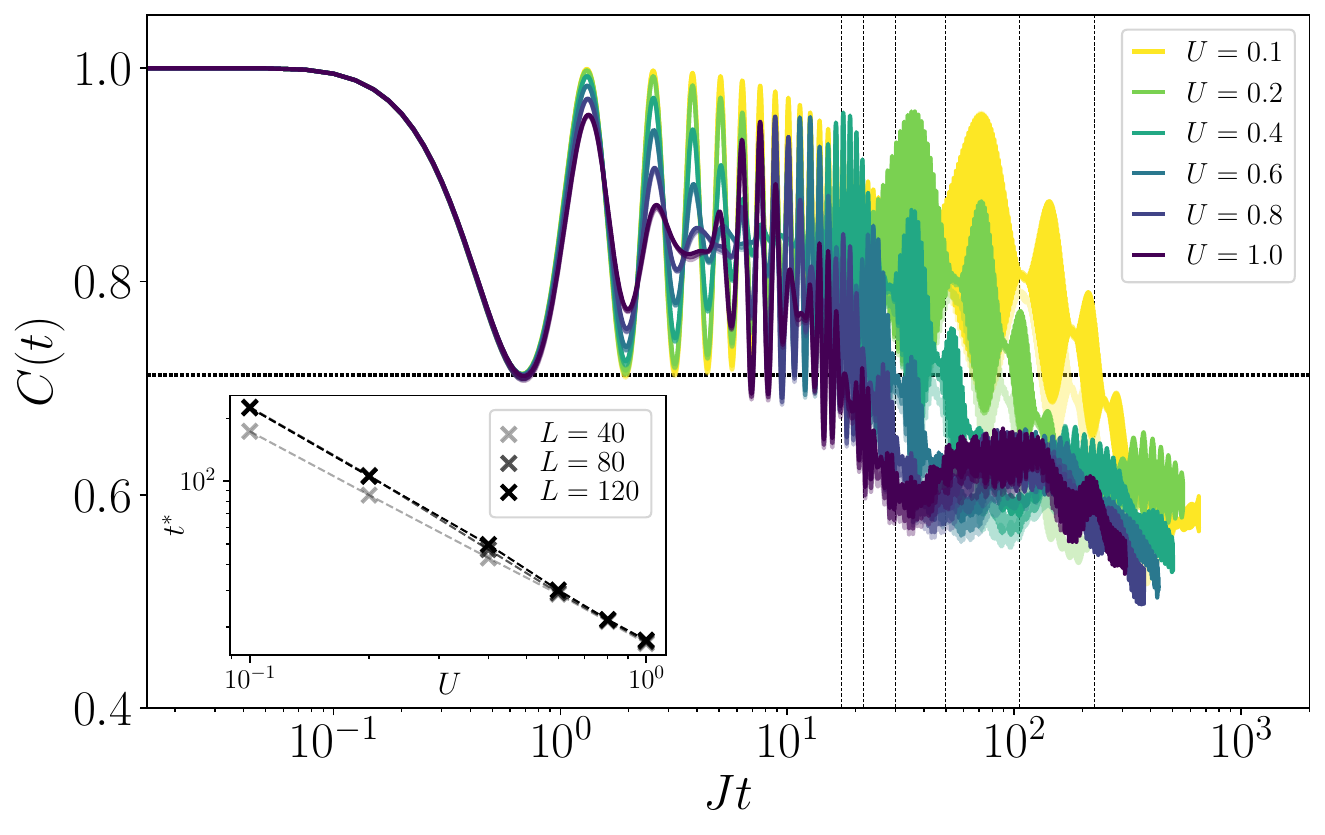}
\caption{tDMRG results for the dynamical autocorrelation function for different interaction strength $U$, $\gamma=5.0$ and $L=40,80,120$. The horizontal line marks the threshold, below which we consider the system to delocalize, given by the amplitude of the first oscillation. Vertical dashed lines mark the delocalization times for different $U$ for $L=120$, obtained with a moving average (see main text). Inset: The delocalization times $t^*$ for different system sizes over $U$.}
\label{fig:stark_1d_dyn_corr_tdvp_deloc_time}
\end{figure}
There are arguments on why a Stark MBL system of finite size has to allow transport at infinite time, since at least a finite set of eigenstates behaves delocalized \cite{Kloss2023}.
 Ref.~\cite{Zisling_2022} numerically investigated the dependency of the resulting delocalization time on system size and found that the delocalization time grows exponentially with system size. This can be argued by introducing a time-dependent gauge transformation, mapping the problem to a prethermal Floquet state, that is exponentially long lived \cite{Abanin_2015, Abanin_2017, Abanin_2017a}. The TFE miss this delocalization effect entirely because it is of higher order than what is included in our order of truncation. 
Since the delocalization time cannot be captured by TFE, which is restricted to small $U$, we suspect that it is a non-perturbative effect.\\
As seen in Fig.~\ref{fig:stark_1d_2d_dyn_corr} the tDMRG method can resolve the delocalization time as long as the bond dimension is large enough, thus, we present tDMRG results here to investigate the dependency of the delocalization time on $U$.  In Fig.~\ref{fig:stark_1d_dyn_corr_tdvp_deloc_time} we show the dynamical autocorrelation computed with tDMRG for different interaction strengths and system sizes. In the localized regime we can simulate larger system sizes. The onset of delocalization is hard to define for the autocorrelation due to strong oscillations. Hence, we establish a heuristic measure by taking the first minimum of the oscillations, i.e. the minimal value in the interval $Jt \in[0,1]$, as the delocalization threshold (horizontal line). The non-interacting system would continue oscillating with this amplitude for $t\to\infty$. Deviations below this threshold, therefore, signify interaction effects. The delocalization time $t^*$ is set to be the crossing of the moving time averaged autocorrelation below this threshold (vertical dashed lines). The averaging is necessary due to persisting high-frequency oscillations. We checked that the chosen interval for the moving average has no qualitative impact on the results as long as it is sufficient to flatten high-frequency oscillations.\\
We find $t^* \sim 1/U$. In conjunction with the fact, that TFE do not see this effect, this shows that the delocalization effect is driven by interaction and is an emergent many-body effect that is challenging to access within perturbative expansions.

\section{Discussion}\label{section: Discussion}
 Using a recently developed numerical variant of Wegner flow equations that leverages GPU hardware we investigated Stark MBL in an unexplored range of system sizes in 1D and 2D. Our results show that Stark MBL has hallmark features that are markedly different to conventional MBL. We find that the transition does not drift to larger potentials with system size but to \textit{smaller} ones, so that one might conclude that there is no true delocalized phase in this model in the thermodynamic limit. We could further show the higher-order nature of the present delocalization at long times and how it is a true many-body effect that depends strongly on interaction strength. The underlying mechanisms of the localization transition and dynamics in Stark MBL systems remain poorly understood and are therefore a very relevant subject of future investigations. Understanding the fundamental differences to conventional MBL could also shed light on open questions there.

The TFE present themselves as a very general tool to study finite-size quantum many-body interacting systems beyond ED. If the system is not localized or displays long-range interactions, TFE 
 can even outperform tDMRG which makes it especially useful for 2D systems which can be treated by simply mapping them to a long-range chain. Therefore, it is a very promising candidate to study other systems, for example the Hubbard model in the weak to moderate interaction regime. The TFE could allow to map out a tentative phase diagram of the 2D Hubbard model up to $10\times10$ sites with broken translational invariance, c.f. Ref.~\cite{Kaczmarek_2023}. A massively parallel implementation on distributed GPUs might even challenge Monte Carlo results \cite{Qin_2016} at small to intermediate $U$. Furthermore, one could simulate more exotic models that exhibit topological order \cite{Laflorencie_2022} or hybrids like the Hubbard-Holstein model \cite{Costa_2020}. A matter of ongoing work is implementing TFE in Floquet space, where one can employ the generator of Ref.~\cite{Thomson_2021}, which would enhance capabilities in Floquet engineering \cite{Decker_2020}, as well as enable studying prethermalization \cite{Weidinger_2017} for system sizes beyond ED and Tensor Networks.

\acknowledgments 
This work was supported by the Deutsche Forschungsgemeinschaft (DFG, German Research Foundation) through the
grant KA 3360/4-1 (project number 508440990). Simulations
were performed with computing resources granted by RWTH
Aachen University under project rwth1543. 

\section*{data availability}
The data that support the findings of this article are not
publicly available. The data are available from the authors
upon reasonable request.

\begin{appendix}
\section{Computing commutators}\label{section: Commutators}
\begin{figure*}[htbp]
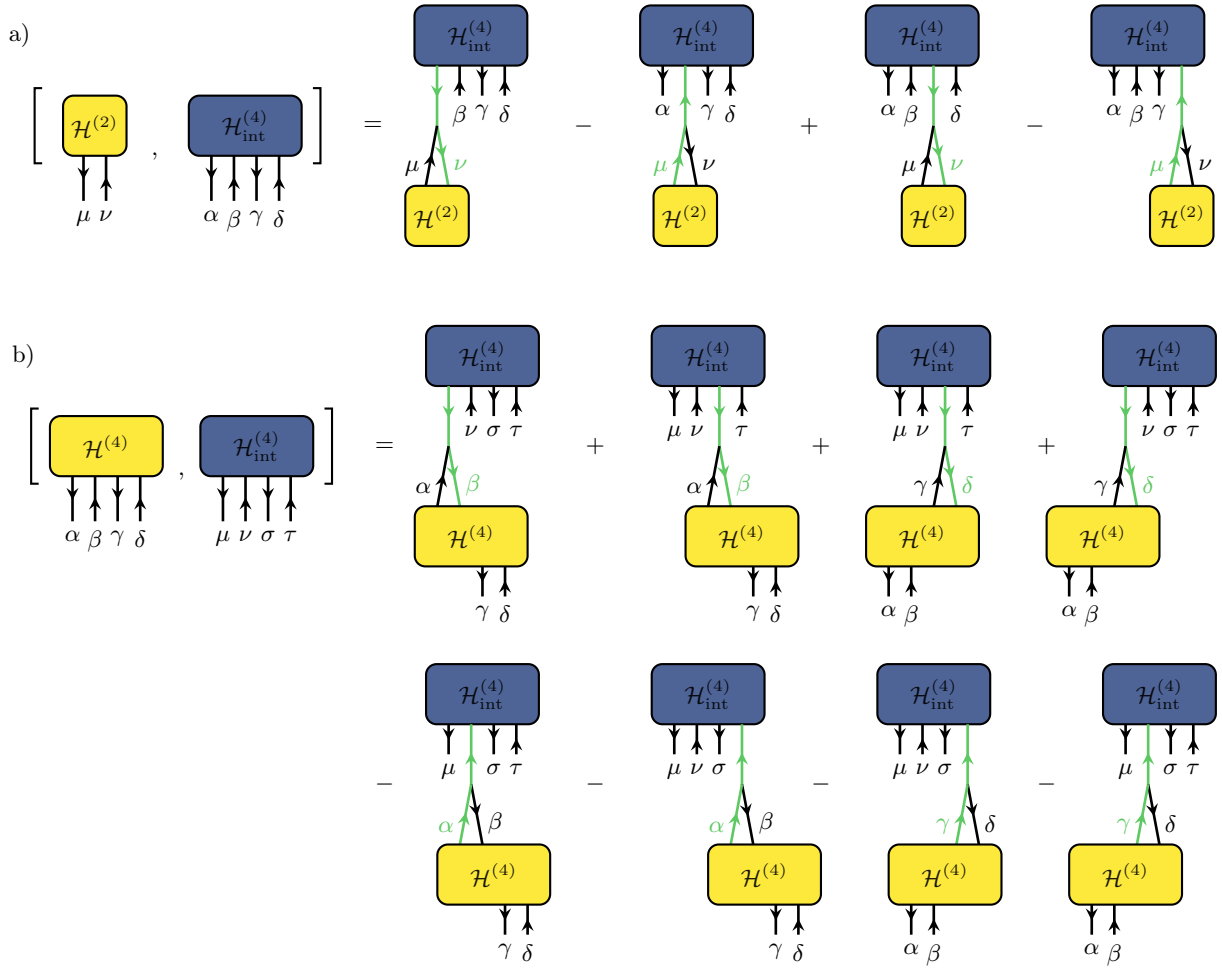

    \centering
    \include{graphical_commutators}
\caption{Graphical representation of computing commutators with a) Four and two legs resulting in four new four-legged tensors, b) Four and four legs, that result in eight new six-legged tensors. To obtain the commutation result the tensors need to be summed with correct sign and index reordering. Contraction rules are given in the main text.}
\label{fig: graphical notation}
\end{figure*}
In order to compute the commutators in Eq.~\eqref{def: Wegner Generator by order}, we follow the procedures that are introduced in Ref.~\cite{Kehrein_2006}. The resulting operator string defines the contraction rule for commuting the coefficient tensors. We will compute the first non-trivial commutator: $\left[:c^{\dagger}_{\mu}c_{\nu}:, :c^{\dagger}_{\alpha}c_{\beta}c^{\dagger}_{\gamma}c_{\delta}: \right]$ as an example (commutators of two legged tensors reduce to the conventional commutator of two matrices with normal ordering corrections). We recall the description of fermionic anticommutation relations in the context of normal ordering:
\begin{equation}
    \langle\{ c^{\dagger}_{\alpha}, c_{\beta}\}\rangle = G_{\alpha \beta} + G_{\beta \alpha}\,,
    \label{def:Anticommutation relation of fermions}
\end{equation}
where $G_{\alpha \beta} = \langle c^{\dagger}_{\alpha} c_{\beta} \rangle$ and $G_{ \beta \alpha} = \langle c_{\beta} c^{\dagger}_{\alpha}  \rangle$ with $\langle ... \rangle$ the expectation value with respect to a deliberately chosen state. Here, we decide to use the vacuum state which implies:
\begin{align}
    &G_{\alpha \beta} =  \langle c^{\dagger}_{\alpha} c_{\beta} \rangle = 0 \,, \label{eq:vacuum normal ordering 1} \\
    &G_{ \beta \alpha} = \langle c_{\beta} c^{\dagger}_{\alpha}  \rangle =  \delta_{\alpha \beta}\,.
\label{eq:vacuum normal ordering 2}
\end{align}
In principle one should use normal ordering with respect to the interacting ground state, but this is difficult to implement in practice \cite{Kehrein_2006}.
Further, we need Wick's second theorem to simplify operator strings:
\begin{equation}
    :O_1::O_2:=:\text{exp}\Bigg(\sum_{k,l} G_{kl} \frac{\partial^2}{\partial A_{k}\partial A'_l}\Bigg)O_1(A)O_2(A'):\Big|_{A'=A},
    \label{def:Wick's second theorem}
\end{equation}
where $A$ is defined as a collection of fermionic operators, operators on the right hand side of Eq.~\eqref{def:Wick's second theorem} can be considered as different sets of fermionic strings. By using Eq.~\eqref{def:Wick's second theorem} we obtain:
\begin{equation}
\begin{split}
    &:c^{\dagger}_{\mu}c_{\nu}: :c^{\dagger}_{\alpha}c_{\beta}c^{\dagger}_{\gamma}c_{\delta}: \\ &=:\Big(1
    +G_{\mu\beta}\frac{\partial^2}{\partial c^{\dagger}_{\mu}\partial c_{\beta}}
    +G_{\mu\delta}\frac{\partial^2}{\partial c^{\dagger}_{\mu}\partial c_{\delta}}
    +G_{\nu\alpha}\frac{\partial^2}{\partial c^{\dagger}_{\nu}\partial c_{\alpha}}\\
    &\hspace{2em}
    +G_{\nu\gamma}\frac{\partial^2}{\partial c^{\dagger}_{\nu}\partial c_{\gamma}}
    +G_{\mu\beta}G_{\nu\alpha}\frac{\partial^4}{\partial c^{\dagger}_{\mu}\partial c_{\beta}\partial c^{\dagger}_{\nu}\partial c_{\alpha}}\\
    &\hspace{2em}
     +G_{\mu\beta}G_{\nu\gamma}\frac{\partial^4}{\partial c^{\dagger}_{\mu}\partial c_{\beta}\partial c^{\dagger}_{\nu}\partial c_{\gamma}}
     +G_{\mu\delta}G_{\nu\alpha}\frac{\partial^4}{\partial c^{\dagger}_{\mu}\partial c_{\delta}\partial c^{\dagger}_{\nu}\partial c_{\alpha}}\\
     &\hspace{2em} +G_{\mu\delta}G_{\nu\gamma}\frac{\partial^4}{\partial c^{\dagger}_{\mu}\partial c_{\delta}\partial c^{\dagger}_{\nu}\partial c_{\gamma}}\Big)c^{\dagger}_{\mu}c_{\nu}c^{\dagger}_{\alpha}c_{\beta}c^{\dagger}_{\gamma}c_{\delta}: \,.\\
\end{split}
\label{def:complicated contraction}
\end{equation}
Therefore, we get the following expression of our typical commutator with normal ordering:
\begin{align}
    %\begin{split}
        &\left[:c^{\dagger}_{\mu}c_{\nu}:, :c^{\dagger}_{\alpha}c_{\beta}c^{\dagger}_{\gamma}c_{\delta}: \right]\nonumber\\
        &=+\underbrace{G_{\mu\beta}}_{0}:c_{\nu}c^{\dagger}_{\alpha}c^{\dagger}_{\gamma}c_{\delta}:-\underbrace{G_{\beta\mu}}_{\delta_{\beta\mu}}:c^{\dagger}_{\alpha}c^{\dagger}_{\gamma}c_{\delta}c_{\nu}:+\underbrace{G_{\mu\delta}}_{0}:c_{\nu}c^{\dagger}_{\alpha}c_{\beta}c^{\dagger}_{\gamma}: \nonumber\\
    &\hspace{1em} - \underbrace{G_{\delta\mu}}_{\delta_{\delta\mu}}:c^{\dagger}_{\alpha}c_{\beta}c^{\dagger}_{\gamma}c_{\nu}:+\underbrace{G_{\nu\alpha}}_{\delta_{\nu\alpha}}:c^{\dagger}_{\mu}c_{\beta}c^{\dagger}_{\gamma}c_{\delta}:-\underbrace{G_{\alpha\nu}}_{0}:c_{\beta}c^{\dagger}_{\gamma}c_{\delta}c^{\dagger}_{\mu}:\nonumber\\
     &\hspace{1em}
     +\underbrace{G_{\nu\gamma}}_{\delta_{\nu\gamma}}:c^{\dagger}_{\mu}c^{\dagger}_{\alpha}c_{\beta}c_{\delta}:-\underbrace{G_{\gamma\nu}}_{0}:c^{\dagger}_{\alpha}c_{\beta}c_{\delta}c^{\dagger}_{\mu}:+\underbrace{G_{\mu\beta}}_{0}G_{\nu\alpha}:c^{\dagger}_{\gamma}c_{\delta}:\nonumber\\
     &\hspace{1em} - \underbrace{G_{\alpha\nu}}_{0}G_{\beta\mu}:c_{\delta}c^{\dagger}_{\gamma}:+\underbrace{G_{\mu\beta}}_{0}G_{\nu\gamma}:c^{\dagger}_{\alpha}c_{\delta}:-\underbrace{G_{\gamma\nu}}_{0}G_{\beta\mu}:c_{\delta}c^{\dagger}_{\alpha}:\nonumber\\
     &\hspace{1em}+\underbrace{G_{\mu\delta}}_{0}G_{\nu\alpha}:c_{\beta}c^{\dagger}_{\gamma}:-\underbrace{G_{\alpha\nu}}_{0}G_{\delta\mu}:c^{\dagger}_{\gamma}c_{\beta}:+\underbrace{G_{\mu\delta}}_{0}G_{\nu\gamma}:c^{\dagger}_{\alpha}c_{\beta}:\nonumber\\
     &\hspace{1em}-\underbrace{G_{\gamma\nu}}_{0}G_{\delta\mu}:c_{\beta}c^{\dagger}_{\alpha}:\nonumber\\
     &=-\delta_{\beta\mu}:c^{\dagger}_{\alpha}c^{\dagger}_{\gamma}c_{\delta}c_{\nu}:-\delta_{\delta\mu}:c^{\dagger}_{\alpha}c_{\beta}c^{\dagger}_{\gamma}c_{\nu}:+\delta_{\nu\alpha}:c^{\dagger}_{\mu}c_{\beta}c^{\dagger}_{\gamma}c_{\delta}:\nonumber\\
     &\hspace{1em} +\delta_{\nu\gamma}:c^{\dagger}_{\mu}c^{\dagger}_{\alpha}c_{\beta}c_{\delta}:\nonumber \\
        &=-\delta_{\beta\mu}:c^{\dagger}_{\alpha}c_{\nu}c^{\dagger}_{\gamma}c_{\delta}:-\delta_{\delta\mu}:c^{\dagger}_{\alpha}c_{\beta}c^{\dagger}_{\gamma}c_{\nu}:+\delta_{\nu\alpha}:c^{\dagger}_{\mu}c_{\beta}c^{\dagger}_{\gamma}c_{\delta}:\nonumber\\
    &\hspace{1em}+\delta_{\nu\gamma}:c^{\dagger}_{\alpha}c_{\beta}c^{\dagger}_{\mu}c_{\delta}:\,.
    %\end{split}
    \label{def:complicated commutator}
\end{align}
In the final step we utilized the fermionic rule of swapping operators where each swap intruduces a minus sign. The commutators are not closed, i.e. they result in longer operator strings and accordingly in coefficient tensors of higher order. While cumbersome, the calculation above can be generalized to any order.
In Ref.~\cite{Thomson2023a}, a graphical representation was introduced to simplify the numerical implementation for calculating these commutators. We introduce our own graphical notation in Fig.~\ref{fig: graphical notation}, that allows to directly read off not only which legs are contracted but also the new index order and the sign. To compute commutators one needs to contract over every possible combination of legs, where only outgoing ($\cc{c}$) and ingoing legs ($c$) can be combined. A contraction over two legs is denoted by \begin{tikzpicture}[baseline={(0,-0.1)}]
\draw[line width =.5pt, color=viridisGreen, postaction={on each segment={mid arrow=viridisGreen}}](-.2,0) -- (.2,0);
\draw[line width =.5pt, color=viridisGreen, postaction={on each segment={mid arrow=viridisGreen}}](.2,0) -- (.4,0);
\end{tikzpicture}. The next step is to rearrange the indices to restore normal ordering (we use vacuum normal ordering here) and to obtain the correct sign of the contraction. Connected arrows pointing in opposite direction signify that the index of the black arrow needs to be swapped into the position left by the green arrow after contraction.  \begin{tikzpicture}[baseline={(0,-0.1)}]
\draw[line width =.5pt,postaction={on each segment={mid arrow=black}}](.0,0) -- (-.2,0);
\draw[line width =.5pt](.0,0) -- (.2,0);
\draw[line width =.5pt,postaction={on each segment={mid arrow=black}}](.2,0) -- (.4,0);
\end{tikzpicture} denotes both a swap, and that the term is multiplied by $(-1)$.  \begin{tikzpicture}[baseline={(0,-0.1)}]
\draw[line width =.5pt,postaction={on each segment={mid arrow=black}}](-.2,0) -- (.2,0);
\draw[line width =.5pt,postaction={on each segment={mid arrow=black}}](.4,0) -- (.0,0);
\end{tikzpicture} denotes a swap with no additional sign change.

\section{Computing Operators}\label{section: Operator flow}
Since TFE is a continuous basis change one can also transform an operator $\mathcal{O}$ into the diagonal basis of $\ham$. The form of the operator in the diagonal basis is especially useful for the description of time-dependent observables. Furthermore, this can be used to study the nature of the performed basis change by investigating the change in the operator structure. Generally, even initially local operators become long-ranged during the flow, however, if the operator stays local the model is likely to be MBL \cite{Singh_2021}. The flow of $\mathcal{O}$ can be performed in parallel by computing $\mathcal{O}(l) = U(l)\mathcal{O}(0)\cc{U}(l)$, where $\mathcal{O}(0)$ can be any operator in the microscopic basis that can be written in terms of the same fermionic operators as the Hamiltonian itself. This amounts to solve
\begin{equation}
    \frac{\d \mathcal{O}(l)}{\d l} = \left[\mathcal{O}(l), \eta(l)\right] \, ,
    \label{eq: operator flow equation}
\end{equation}
where the generator $\eta(l)$ is the one constructed from $\ham(l)$. In this work we study the microscopic number operator on site $i$, $\mathcal{O}(0) = n_i = \cc{c}_ic_i$. To account for the generation of non-local elements and even interaction elements during the flow, the operator is initialized in the same format as the Hamiltonian, with only the $i$th on-site term being non-zero. The flow results in a transformed number operator of the form 
\begin{align}
    n_i =& \sum_{\alpha\beta} A_{\alpha\beta} :\cc{\tilde{c}}_{\alpha}\tilde{c}_{\beta}:\nonumber \\ 
    &+ \sum_{\alpha\beta\gamma\delta} B_{\alpha\beta\gamma\delta} :\cc{\tilde{c}}_{\alpha}\tilde{c}_{\beta}\cc{\tilde{c}}_{\gamma}\tilde{c}_{\delta}:\, ,
    \label{eq: Flowed Number operator}
\end{align}
expanded in terms of fermionic operators $\cc{\tilde{c}},\,\tilde{c}$ in the diagonal basis of $\ham$. The operator itself is now neither diagonal nor local anymore, but expectation values with respect to eigenstates become trivial, since they can be represented as simple product states in the diagonal basis. In the main text we drop the tilde and it is clear from context when we are considering operators in the diagonal basis.
\section{Error Analysis} \label{section: energy error analysis}
\begin{figure}
\includegraphics[width=1\columnwidth]{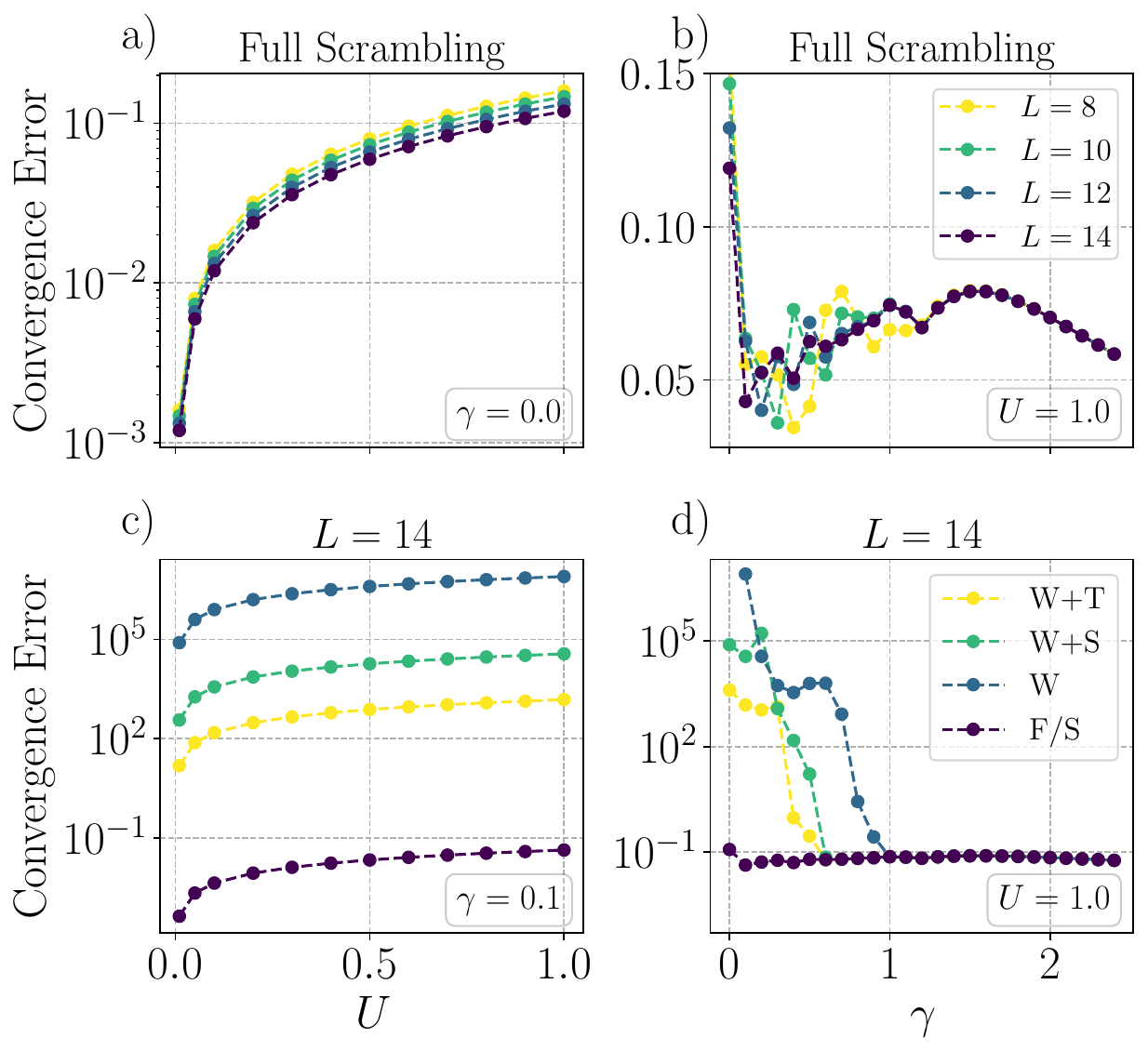}
\caption{The convergence error given by $\mathrm{max}|\ham^{(4)}_{\mathrm{off}}|$ with different choices of generators for the Stark MBL model with varying system sizes. a)-b) show the best choice of generators as we found it for increasing system size and different interaction strength $U$, which is the full scrambling transformation. c)-d) show a system with $L=14$ for the different variants of generator combinations.
''W'' denotes application of the bare Wegner generator, ''W+T'' denotes the combination of Wegner and Toda generator. 
"F/S" denotes a full scrambling transformation before the start of a conventional Wegner flow, threshold $\varepsilon=0.01$. ''W+S'' stands for a combination of Wegner and an infinitesimal scrambling transform with $\varepsilon=0.5$.
}
\label{fig:convergence_error_stark_mbl}
\end{figure}
\begin{figure}[htbp]
\includegraphics[width=1\columnwidth]{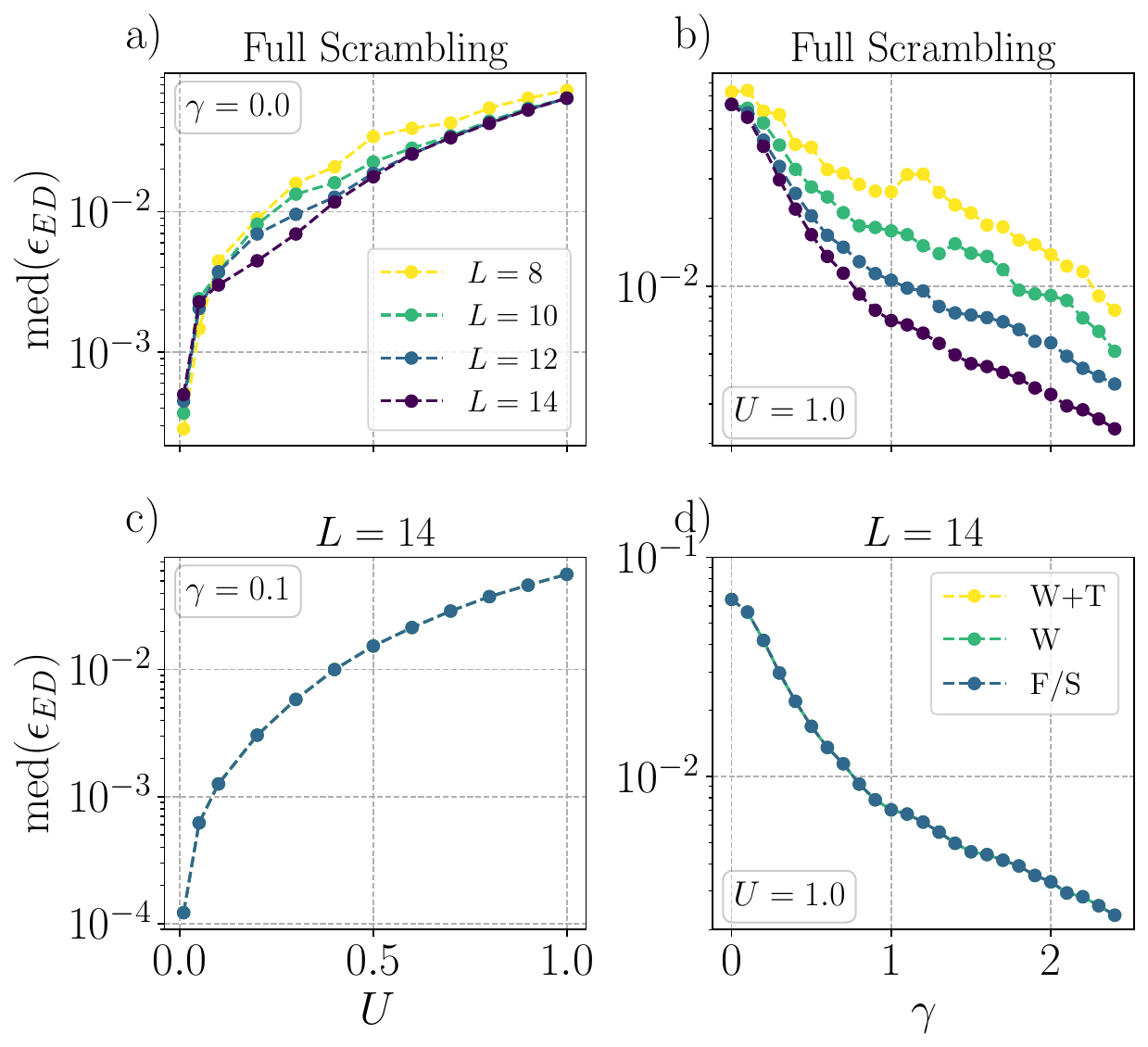}
\caption{The median energy error compared to ED for the Stark MBL model with varying system sizes, interaction strengths, potential strengths and different generators. a)-b) show the energy error for different system sizes with the full scrambling transformation employed at the start of the flow. {c)-d)} show the energy error for different variants of generators tested (Legend is the same as in Fig.~\ref{fig:convergence_error_stark_mbl}). In panels c)-d) all values lie on top of each other. The Wegner generator does not work at $\gamma=0$, therefore we omitted this data point.}
\label{fig:energy_error_stark_mbl}
\end{figure}
\begin{figure}[htbp]
\includegraphics[width=0.8\columnwidth]{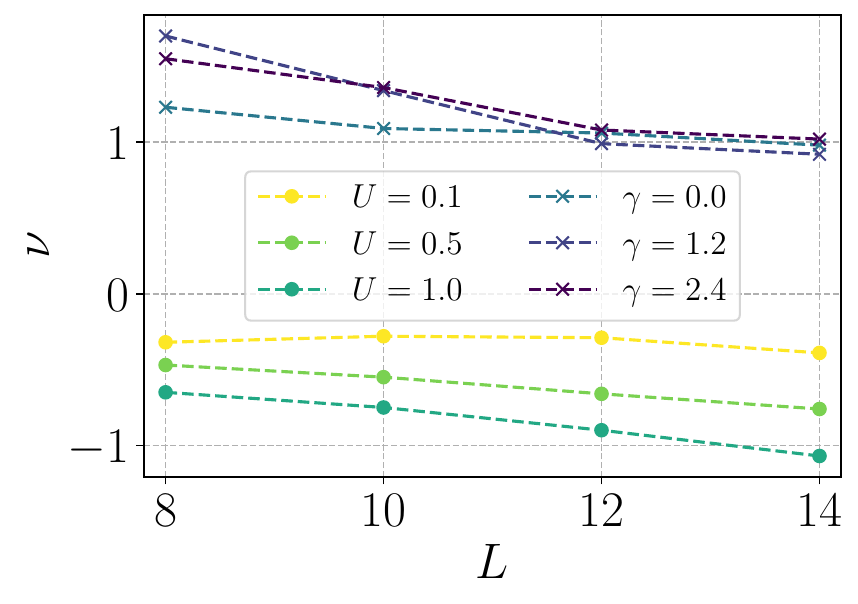}
\caption{A power-law fit was performed to extract the scaling of the energy error with both $\gamma$ and $U$. The underlying data is the same as in Fig.~\ref{fig:energy_error_stark_mbl} a) and b). Dots: The exponent of the scaling with $\gamma$ for different system sizes and interactions, $\epsilon_{ED}(\gamma) \sim \gamma^{\nu_{\gamma}}$. 
Crosses: The exponent of the scaling with $U$ for different system sizes and potential strengths, $\epsilon_{ED}(U) \sim U^{\nu_{U}}$. 
}
\label{fig:energy_error_powerlaw_fit}
\end{figure}

We choose to include operators up to 4th order in the flow equations. This means that we are working with a resummed perturbative expansion, motivated from small interaction strengths. The neglected terms are on the order $\mathcal{O}(U^2/\tilde{\gamma})$, which determines the maximal energy resolution of the approximation \cite{Thomson_2024}. $\tilde{\gamma}$ denotes the effective energy separation between sites, which in the limit of $\gamma\to 0$ is artificially increased by the scrambling transformation. We find no significant improvement in accuracy by including 6th order while encountering significantly increased computational cost.\\ 
In practice we also introduce a convergence error due to stopping the flow at large but finite $l$. We define the convergence error as the maximal remaining off-diagonal element of the interaction terms. This is shown in Fig.~\ref{fig:convergence_error_stark_mbl} for a range of parameters and different choices of generators.  $\mathrm{max}|\ham^{(4)}_{\mathrm{off}}|$ is generally smaller than $U^2$ in magnitude for the systems analysed here. The default Wegner generator performs worst for small $\gamma$ and for $\gamma=0$ the flow does not start at all, as the generator vanishes due to degeneracies.
Panels b) and d) show how the effective energy separation induced by the scrambling transformation significantly improves the results for small $\gamma \lesssim 1.0$. The onset of the scrambling transformation can be modified by the threshold $\varepsilon$ where a smaller value shifts the onset to larger $\gamma$. The choice of $\varepsilon$ does not lead to qualitative changes for $\varepsilon\lesssim 0.5$, however, we found that choosing a small threshold $\varepsilon=0.01$ and being more thorough in lifting degeneracies leads to a more stable convergence behavior. This is particularly the case for 2D systems. While this does not affect the diagonalized Hamiltonian it reduces instabilities in the operator flow. The \textit{full} scrambling transformation is the only one able to diagonalize even at zero external field, i.e.~degenerate single particle states. But, the combination of Wegner generator and Toda generator still allows to extend the accessible parameter range of the canonical flow and at increased speed in the interval $0.6 \lesssim \gamma \lesssim 1.0\,$.\\ 
The convergence error propagates into the error of the eigenenergies that can be extracted from the Hamiltonian after the flow. In Fig.~\ref{fig:energy_error_stark_mbl} we show the median energy error corresponding to state $\ket{j}$, $\mathrm{med}(\epsilon_{ED}) \equiv \underset{j}{\mathrm{med}}(\epsilon^{(j)}_{ED})$, where we defined
\begin{equation}
   \epsilon^{(j)}_{ED} = \left|\frac{E^{(j)}_{TFE} - E^{(j)}_{ED}}{E^{(j)}_{ED}} \right| \,, \label{eq: TFE_ED energy error}
\end{equation}
with respect to results from ED.  A power-law fit, shown in Fig.~\ref{fig:energy_error_powerlaw_fit}, shows that the error in eigenenergies scales qualitatively with $U/\gamma$ in the ED accessible range of system sizes. Fig.~\ref{fig:energy_error_stark_mbl}c) and d) suggest that the chosen generator does not have an impact on the energy error. However, as shown in Fig.~\ref{fig:convergence_error_stark_mbl}d), the 4th order elements only converge with the full scrambling transformation in the limit of $\gamma\to 0$. Only with the converged results we can expect interaction driven effects to be incorporated accurately. To obtain results in Section~\ref{section: Results} we always use the full scrambling transformation.
\begin{figure}[htbp]
\includegraphics[width=1\linewidth]{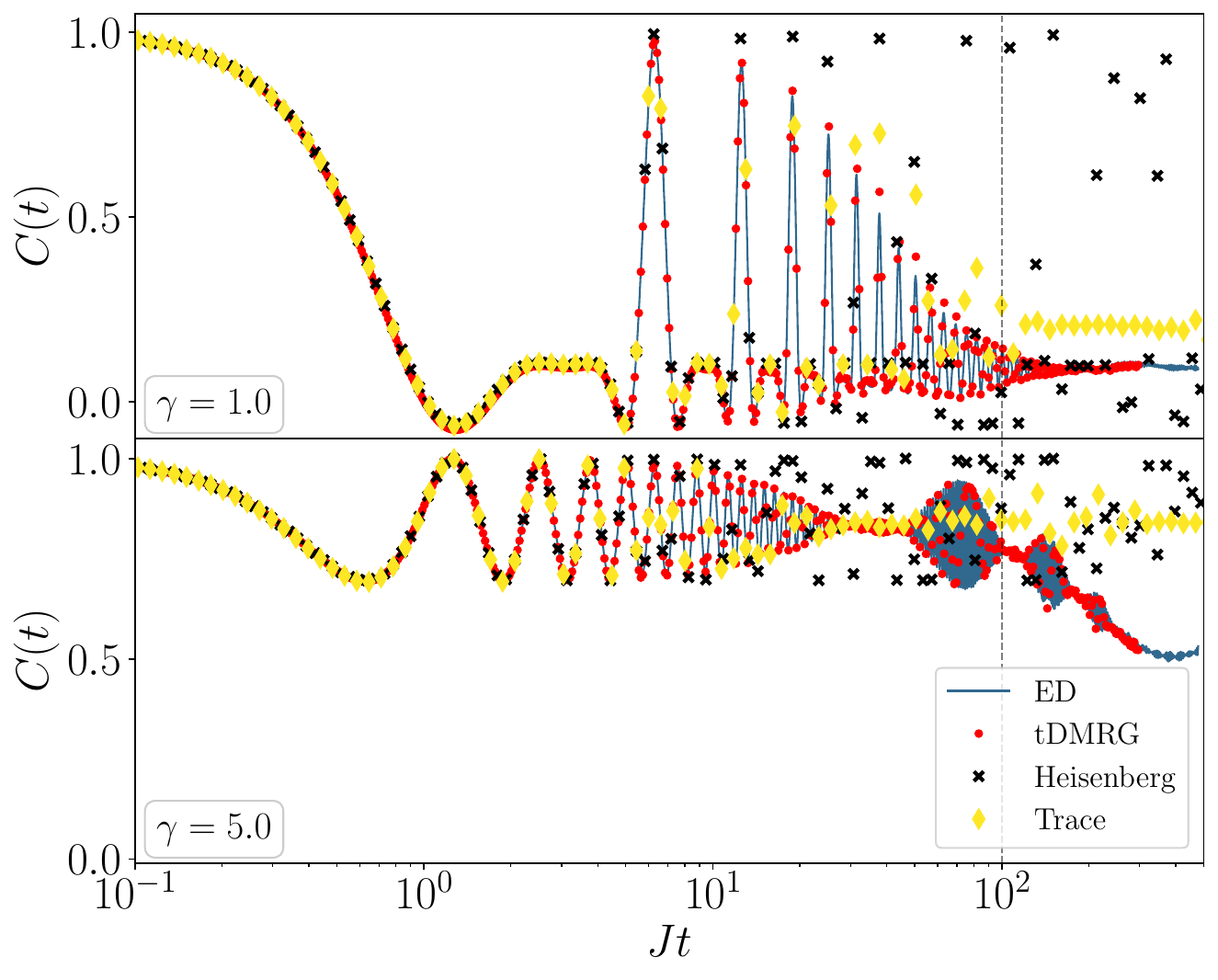}
\caption{The dynamical autocorrelation for a 1D Stark MBL system, system size is $L=14$. Top panel shows weak potential $\gamma=1.0$ while bottom panel shows strong potential $\gamma=5.0$. Interaction strength is set to $U=0.1$. Both the method of calculating the time evolution with the Heisenberg equation of motion (black crosses) and the infinite temperature trace in Eq.~\eqref{eq: Dynamical Autocorrelation from Eigenenergies} (yellow diamonds) are shown. As benchmark, ED and tDMRG data is shown (tDMRG until bond dimension grows too large). TFE accuracy breaks down at times $t \sim 1/U^2$ (dashed line).}
\label{fig:stark_1d_dyn_corr_comparison}
\end{figure}

\section{ Comparison of Dynamics Implementations}\label{section: Dynamics comparison}
Ref.~\cite{Thomson_2024} computed Eq.~\eqref{eq:dynamical_autocorr} in a different manner for disordered and QP-MBL systems. Instead of first computing the energies and implementing time evolution as a phase and then summing over all operator matrix elements, one can also solve the Heisenberg equations of motion for $n_i(t)$ in the diagonal basis. Then, one obtains the correlator by multiplying the solution of the Heisenberg equation with the number operator at $t=0$, and averaging the expectation values of randomly drawn eigenstates. 
\begin{align}
\langle n_i(t)n_i(0) \rangle &= \langle e^{i\ham t}n_ie^{-i\ham t}n_i\rangle \nonumber\\
&= \langle \int_0^t i\left[\ham,n_i(t')\right]\d t' \cdot n_i(0) \rangle \,. \label{eq: Dynamical Autocorrelation from Heisenberg EOM}
\end{align}
We compared both ways to extract $C(t)$ and find that the truncation error induced by the flow is larger when solving the Heisenberg equation. In Fig.~\ref{fig:stark_1d_dyn_corr_comparison} we present the resulting dynamics from both Eq.~\eqref{eq: Dynamical Autocorrelation from Eigenenergies} and Eq.~\eqref{eq: Dynamical Autocorrelation from Heisenberg EOM} in comparison to ED and tDMRG. With the Heisenberg equation of motion, accurate simulations without disorder averaging can only be achieved up to times $t\sim 1/U$, severely underestimating the damping effects introduced by interaction. In models with random disorder these errors average out after taking the mean over several disorder configurations. However, we need to go beyond that to make the TFE viable for clean systems. Using Eq.~\eqref{eq: Dynamical Autocorrelation from Eigenenergies} we get accurate results up to $t \sim 1/U^2$ as we accurately capture the damping effects at intermediate times. Thus, we can simulate disorder-free models up to times where energy scales smaller than the truncation error become relevant. Importantly, this does not require the system to be localized.
\end{appendix}

%\bibliographystyle{apsrev4-2}
%\bibliography{ref}

\end{document}